\documentclass[sn-mathphys]{sn-jnl}

\jyear{2021}%

\theoremstyle{thmstyleone}%
\theoremstyle{thmstyletwo}%

\theoremstyle{thmstylethree}%

\usepackage{amsmath}
\usepackage{siunitx}
\usepackage{booktabs}
\newcommand{\adiff}{a_{\mathrm{diff}}}
\newcommand{\aspec}{a_{\mathrm{spec}}}
\newcommand{\mco}{m_{\mathrm{co}}}

\begin{document}

\title[Optimizing Multi-Spacecraft Cislunar SDA
Via Hidden-Genes Genetic Algorithm]{Optimizing Multi-Spacecraft Cislunar Space Domain Awareness Systems
via Hidden-Genes Genetic Algorithm}


\author[1]{\fnm{Lois} \sur{Visonneau}}\email{lvisonneau3@gatech.edu}
\equalcont{These authors contributed equally to this work.}

\author[1]{\fnm{Yuri} \sur{Shimane}}\email{yuri.shimane@gatech.edu}
\equalcont{These authors contributed equally to this work.}

\author*[1]{\fnm{Koki} \sur{Ho}}\email{kokiho@gatech.edu}

\affil*[1]{\orgdiv{Daniel Guggenheim School of Aerospace Engineering}, \orgname{Georgia Institute of Technology}, \orgaddress{\street{620 Cherry St NW}, \city{Atlanta}, \postcode{30332}, \state{GA}, \country{USA}}}


\abstract{This paper proposes an optimization problem formulation to tackle the challenges of cislunar Space Domain Awareness (SDA) through multi-spacecraft monitoring. Due to the large volume of interest as well as the richness of the dynamical environment, traditional design approaches for Earth-based architectures are known to have challenges in meeting design requirements for the cislunar SDA; thus, there is a growing need to have a multi-spacecraft system in cislunar orbits for SDA.
The design of multi-spacecraft-based cislunar SDA architecture results in a complex multi-objective optimization problem, where parameters such as number of spacecraft, observability, and orbit stability must be taken into account simultaneously. Through the use of a multi-objective hidden genes genetic algorithm, this study explores the entirety of the design space associated with the cislunar SDA problem. 
A demonstration case study shows that our approach can provide architectures optimized for both cost and effectiveness.}




\maketitle

\section{Introduction}\label{sec1}
The cislunar space is gaining traction as increasing numbers of both human and robotic exploration missions are being planned and launched in the coming years. 
With greater traffic in the cislunar space, the need for space domain awareness (SDA) extended further out of the Earth-orbit region is primordial. 
While the Low-Earth Orbit (LEO), Medium Earth Orbit (MEO), and Geosynchronous/Geostationary Earth Orbit (GSO/GEO) regimes are well-understood and regulated through existing infrastructures, these existing techniques are not always extendable to the cislunar domain \cite{Holzinger2021_report}; one primary contributor to this difficulty is the nature of the dynamics significantly affected by both the Earth and the Moon's gravitational forces; the motion of objects in the three-body problem, even in the case of simplified models, are well-known to be chaotic \cite{Frueh2021}. 
From the perspective of sensor tasking, the challenge also rise from the vast volume that a cislunar SDA architecture must cover compared to an Earth-orbit SDA architecture. 
In terms of orders of magnitude, the Earth-Moon distance is roughly 10 times the GEO altitude, while low-energy transfers may have apogees extending to around 1.5 million km, roughly five times the Earth-Moon distance. Traditional Earth-based sensors have difficulty covering such a large volume, particularly the region near the direction of the Moon; rather, spacecraft tasked with performing observations of targets need to be placed into appropriate orbits. 

Several previous studies proposed approaches for studying architectures tasked for cislunar SDA. 
While earlier works have concentrated on the coverage of the lunar surface \cite{Grebow2008}, the past few years saw a rapid growth in the number of works focusing on detecting, tracking, and maintaining custody of both cooperative and non-cooperative assets in the cislunar region \cite{Fowler2020,Cunio2020,Bolden2020,Dao2020,Thompson2021,Frueh2021b,Vendl2021,Wilmer2022,Dahlke2022,Fedeler2022,Klonowski2022}. 
The vastness of the cislunar space necessitates the use of orbits that extend beyond the region between LEO and GEO, where traditional surveillance and tracking architectures currently reside. 
The region \textit{outside} GEO, commonly referred to as xGEO in the US Defense sector, typically results in two types of orbital locations that may be used: the prior are highly elliptic Earth orbits that exhibit resonance with the Earth-Moon system, and the latter are libration point orbits (LPOs) that revolve around any of the five Lagrange points of the Earth-Moon system. 

Some variation in the definition of the cislunar ``domain'' of interest is seen in literature, and this has a direct impact on the type of orbit(s) that may be useful. 
For example, the works by Wilmer et al \cite{Wilmer2022} and Dahlke et al \cite{Dahlke2022} study SDA activities for the vicinity of Earth-Moon L1, L2, or L3 points; hence, these are done by employing LPOs about these points. 
In contrast, there are also numerous studies considering vast, cislunar \textit{regions} of interest. 
For example, Vendl and Holzinger \cite{Vendl2021} considered a trapezoidal region of interest lying in the Earth-Moon orbital plane, extending from GEO to the far side of the Moon, in this case, the authors have considered the use of LPO families in the vicinity of L1 and L2. 
An even larger area of interest is considered in the works of Cunio et al \cite{Cunio2020}, Bolden et al \cite{Bolden2020} and Frueh et al \cite{Frueh2021b}, where the entire region engulfed by the orbit of the Moon is to be monitored. In this case, Earth-Moon resonant orbits are promising candidates as their tracks span over large volumes of the cislunar space. 
Some other works have considered simulating the tracking performance of actual, flown, planned, or hypothetical translunar trajectories or cislunar spacecraft. 
The works of Dao et al \cite{Dao2020}, Fowler et al \cite{Fowler2020}, and Thompson et al \cite{Thompson2021} combine sensor models with candidate observer orbits to conduct a high-fidelity simulation of tracking cislunar objects. 

For cislunar SDA scenarios involving coverage of regions in the cislunar space, observation limitations using single spacecraft have been identified \cite{Vendl2021}. 
Recent work \cite{Badura2022} has shown the importance of having sufficient geometric diversity between multiple observers in order to obtain high observability. 
Nonetheless, questions regarding the optimal number and distribution of observers in space still remain unanswered.

This paper's focus is on optimizing the multi-satellite architecture for cislunar SDA. Optimizing the design of such an architecture that ensures performance while also minimizing the cost involved in its maintenance is a challenging task, where a varying number of spacecraft must be considered online.
In particular, the main complexity in choosing the appropriate architecture is in the combinatorial nature of the problem, as the number of spacecraft considered and the choice of orbits have infinite possibilities.
%
%

To allow for a variable number of observers trading off the overall architecture cost and effectiveness, an algorithm capable of handling a variable-length decision vector must be considered. 
In this work, a multi-objective variant of the hidden-genes genetic algorithm (HGGA) \cite{Gad2011,Abdelkhalik2016,Abdelkhalik2018} is employed. HGGA allows for the exploration of the entire design space in one search, simultaneously varying the continuous variables, the periodic orbits periods and  phases angles, as well as the discrete variables such as the choice of orbits considered and the number of spacecraft required. 
Through the use of the multi-objective HGGA, we are able to obtain Pareto front solutions corresponding to a variety of cislunar SDA architectures, providing insights into the trade-offs between performance and costs for this problem. 

This paper is organized as follows: firstly, Section \ref{sec:simulator} introduces the simulation environment, including the dynamical system, definitions of the target distribution, and measures to quantify the performance of a cislunar SDA architecture. 
Then, in Section \ref{sec:optim_hgga}, the multiobjective optimization problem formulation, along with the multi-objective HGGA algorithm, are introduced.
This is followed by Section \ref{sec:numerical_results}, where the proposed formulation is employed to obtain Pareto front solutions of cislunar SDA architectures. 
Finally, Section \ref{sec:conclusion} concludes this work.

\section{Simulation of Cislunar Space Domain Awareness Architecture}
\label{sec:simulator}
The problem of cislunar space domain awareness requires the development of a dedicated simulation environment that tracks the location of the Earth, Moon, Sun, observer(s), and target(s). 
To this end, the Earth-Moon circular restricted three-body problem (CR3BP), along with the associated Earth-Moon rotating frame, is particularly advantageous as it provides a reference frame in which the Earth and Moon may be assumed to be constant. 
In this work, the targets are considered to be stationary points distributed in cislunar space, while

\subsection{Dynamical System}
The study of motion in the Earth-Moon system using the CR3BP has been highly successful due to the relatively low eccentricity of the Moon's orbit and the small effect of the Sun's gravitational effect, especially in the vicinity of the Earth or Moon \cite{Frueh2021}. 
The CR3BP equations of motion are given by
\begin{equation}
\begin{aligned}
    \ddot{x} - 2\dot{y} &= \dfrac{\partial U}{\partial x}
    \\
    \ddot{y} + 2\dot{x} &= \dfrac{\partial U}{\partial y}
    \\
    \ddot{z} &= \dfrac{\partial U}{\partial z}
\end{aligned}
\label{eq:eom_CR3BP}
\end{equation}
where $U$ is the pseudo-potential given by
\begin{equation}
    U = \dfrac{x^2 + y^2}{2} + \dfrac{1-\mu}{r_1} + \dfrac{\mu}{r_2}
    \label{eq:cr3bp_pseudoPotential}
\end{equation}
and $\mu$ is the mass-parameter of the CR3BP system, given by
\begin{equation}
    \mu = \dfrac{{M}_2}{M_1 + M_2}
\end{equation}
where {$M_1$} is the mass of the primary body, and {$M_2$} is the mass of the secondary body. 
The corresponding state-transition matrix (STM) can be propagated along with the states through the initial value problem
\begin{equation}
    \left\{\begin{array}{l}
        \dot{\Phi} (t) = \boldsymbol{A} (\boldsymbol{x}) \Phi(t)
        \\
        \Phi(0) = \mathbf{I}_{6 \times 6}
    \end{array}\right.
\end{equation}
where $\boldsymbol{A}$ is the Jacobian of the dynamics. 

\subsection{Periodic Orbit and Database}
Periodic orbits about libration points of the CR3BP (LPOs) are promising candidates for cislunar SDA architectures due to their positioning within the cislunar space. Specifically, the L1 and L2 Lyapunov and halo families, as well as the distant retrograde orbit (DRO) family offer both favorable geometric coverage within the space of interest and are known to have acceptable or high stability properties suitable for long-term use. 

A periodic orbit must obey the relation 
\begin{equation}
    \boldsymbol{x}(t + P) = \boldsymbol{x} (t)
\end{equation}
where $\boldsymbol{x}$ are the states, and $P$ is the period. In the context of the LPOs of the selected families, they are known to exhibit a symmetry about the $xz$-plane in the Earth-Moon rotating plane; leveraging this result, the well-known single-shooting scheme \cite{Howell1984}
\begin{equation}
    \begin{aligned}
        & X_{k+1} = X_{k} - \left[\dfrac{\partial F(X_k)}{\partial X}\right]^{-1} F(X_k)
    \end{aligned}
\end{equation}
where the residual $F$ is defined by
\begin{equation}
    F = \begin{bmatrix}
            y(P/2) \\ \dot{x}(P/2) \\ \dot{z}(P/2) 
        \end{bmatrix}
\end{equation}
and the design vector $X$ contain 3 out of the 4 free variables, namely $x(0)$, $z(0)$, $\dot{y}(0)$, and the period $P$, can be employed. The Jacobian $\dfrac{\partial F(X_k)}{\partial X}$ can be simply constructed from combinations of the entries of the STM and the equations of motion if $T$ is included in $X$. 
The choice of variables to be included in $X$ is application dependent; in the context of creating a database of LPOs, the choice is made based on the ease of convergence to the LPO in the desired range of $X$. For example, low-amplitude halos are difficult to obtain unless $z(0)$ is fixed to a desired non-zero value. 

Once an LPO is obtained, its linear stability index $\nu$ can be defined by evaluating the eigenvalues of its monodromy matrix, which corresponds to the STM at $t=P$. Specifically, $\nu$ is given by \cite{Grebow2008}
\begin{equation}
    \nu = \dfrac{1}{2} \left( \lambda_{\max} + \dfrac{1}{\lambda_{\max}} \right)
\end{equation}
where $\lambda_{\max}$ is the largest eigenvalue of $\Phi(P)$.

\begin{figure}[ht!]
    \centering
    \includegraphics[width=0.55\textwidth]{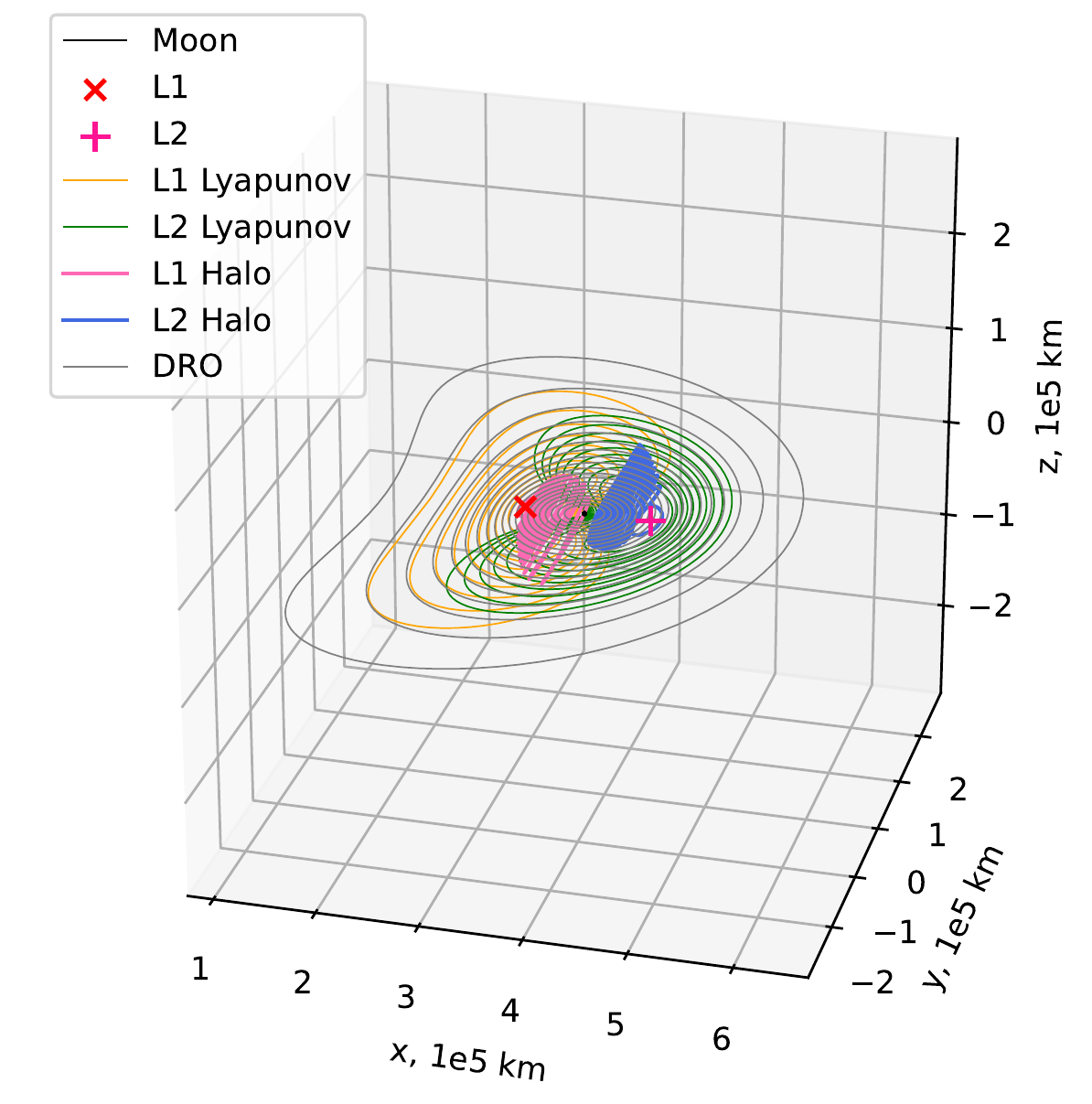}
    \caption{Example Libration Point Orbits for the Earth-Moon system studied in this work}
    \label{fig:lpo_families}
\end{figure}

A finite number of periodic orbits belonging to the selected families are first calculated using a single-shooting method. Figure \ref{fig:lpo_families} shows a sample set of the LPOs constructed for this work. 
This results in a database that can be interpolated against properties such as the $x$-axis crossing points on the $xz$-plane, or the periods, of each LPO. 
This database can then be linearly interpolated to yield an initial guess of an arbitrary $x$-axis crossing or period, which can then rapidly converge to the corresponding LPO. 

When considering the location of a spacecraft in LPOs, its position along the orbit at a given time must also be considered due to the revolution of the Sun. We define an initial phase angle $\phi \in [0,2\pi]$ such that the spacecraft's initial position along an LPO is given by
\begin{equation}
    \boldsymbol{x}(t_0) = \boldsymbol{x}\left( \dfrac{\phi}{2\pi} P \right)
\end{equation}

\subsection{Sun-Direction}
Along with the state of the observer spacecraft and the location of the observing targets, the location of the Sun is a fundamental quantity that must be identified when simulating SDA activities. In the context of trajectory design, the bi-circular restricted four-body problem (BCR4BP), which considers the Sun's motion to be along a circular path on the $xy$-plane of the Earth-Moon rotating frame, is an insightful model to generate transfers that leverage gravitational effects of all three bodies, such as low-energy transfers \cite{Koon2001,belbruno2002analytic,jpl-monograph-series12,Boudad2020}. 
However, for the purpose of analyzing SDA architectures, it is important to consider the offset between the Earth-Moon and Sun-Earth planes. 
As such, in this work, JPL NAIF's SPICE data is used to obtain the vector from the Earth-Moon barycenter to the Sun. 
During the simulation, an arbitrary reference epoch is chosen, and this Sun-direction vector is extracted at each time step where the observation is to be simulated. 

\begin{figure}
    \centering
    \includegraphics[width=0.99\linewidth]{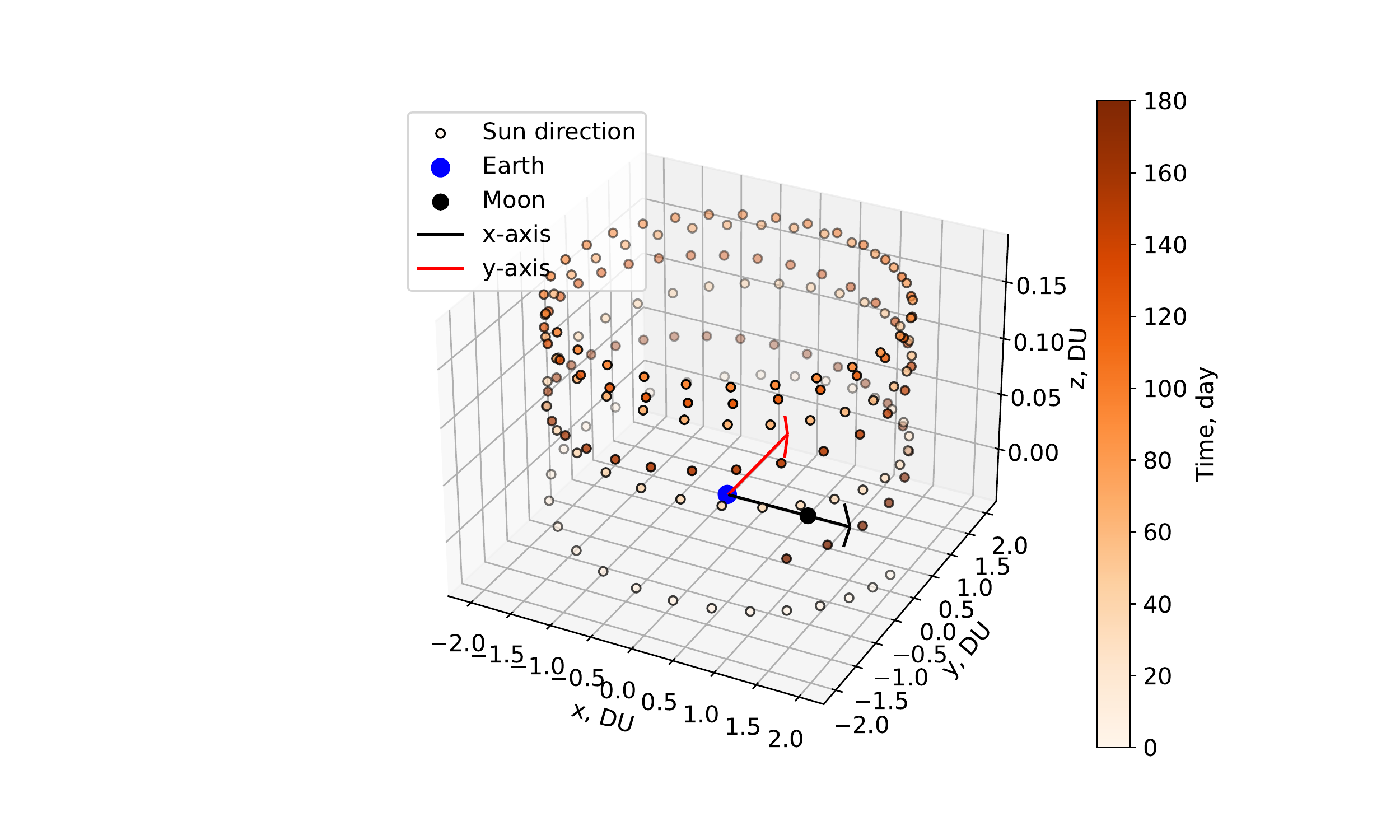}
    \caption{Sun directions over 180 days, in the Earth-Moon rotating frame}
    \label{fig:sun_dirs}
\end{figure}

As an example, the Sun-direction is plotted for a 180 days period in the Earth-Moon rotating frame in Figure \ref{fig:sun_dirs}, starting on 2024 MAR 25 07:12:36.633. 
This reference epoch has been chosen simply as a future date where the Sun lies on the $xz$-plane of the Earth-Moon rotating frame. 
Note that the scattered circles do not represent the actual position of the Sun, but rather its direction in the Earth-Moon rotating frame. 
It is possible to observe the Sun's helical path, which results in a variation of the direction of the illuminating source in cislunar space. 

\subsection{Observation Targets and Metric}
Comparing different cislunar architectures requires the definition of metrics. Based on \cite{Vendl2021}, we consider an observation metric allowing for the comparison of the persistent detection of the cislunar space for a given mission horizon. 
To this end, we define a spatial observation zone as global as possible so that any trajectory of a spacecraft leaving earth for the vicinity of the moon or the surrounding would have its trajectory entirely included in the volume defined. Within this zone, a discrete number of targets are placed in a gridded fashion. 
Then, based on the locations of the observer(s), targets, and the Sun, the \textit{observability metric} $\Gamma$ is defined as a scalar value that quantifies the viability of a given SDA architecture.

\subsubsection{Target Distribution}
\begin{figure}
    \centering
    \includegraphics[width=0.5\linewidth]{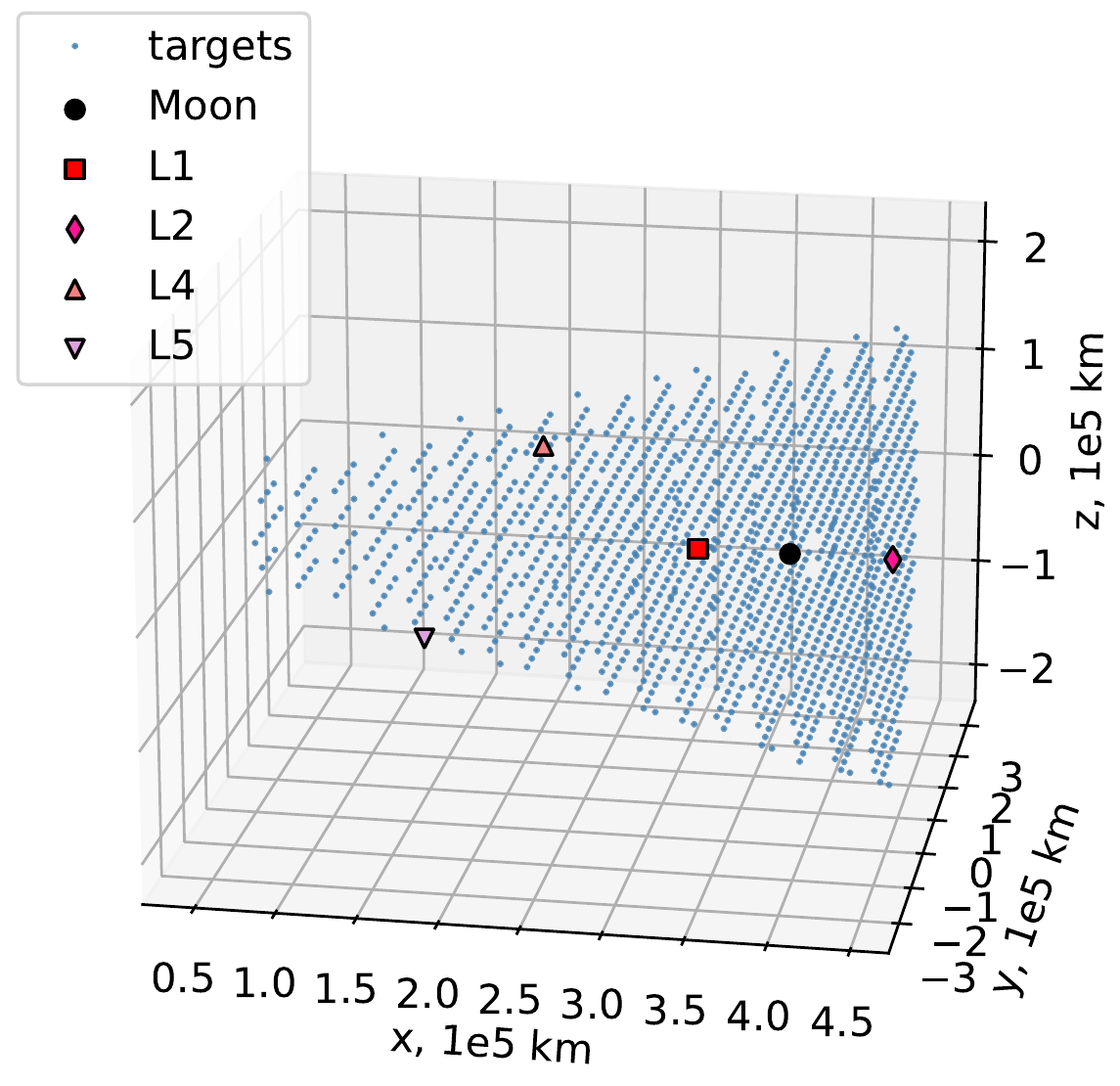}
    \caption{Equispaced targets distributed in cislunar space in the Earth-Moon rotating frame}
    \label{fig:targets}
\end{figure}

A set of static targets that encompasses the cislunar space are considered in this work. The targets are equispaced in Cartesian space within a truncated cone of interest that encompasses the region between the GEO belt and Earth-Moon L2. 
The choice of this set of targets is motivated by the so-called ``cone of shame'', which is the lunar exclusion zone identified by the Air Force Research Laboratory to be an area of high interest for monitoring \cite{afrl2020}. 
Targets used in this work are shown in Figure \ref{fig:targets}. 

\subsubsection{Observation Metric}
The visibility of an object is measured by the apparent magnitude, which is a relative scale for quantifying the brightness of objects. 
This work follows the model used by Vendl and Holzinger \cite{Vendl2021}, which is presented here for completeness. 
Suppose the target has a diameter $d$ and is at a range $\zeta$ from the observer. 
The apparent magnitude of a target is given by
\begin{equation}
    m_{\mathrm{target}} = m_{\mathrm{S}} - 2.5 \log_{10} \left( \dfrac{d^2}{\zeta^2} \left[ 
        \dfrac{a_{\mathrm{spec}}}{4} + a_{\mathrm{diff}} p_{\mathrm{diff}} (\psi)
    \right] \right)
\end{equation}
where $m_{\mathrm{S}} = -26.74$ is the apparent magnitude of the Sun, $\psi$ is the solar phase angle given by
\begin{equation}
    \psi = \arccos{
    \left(
        \dfrac{
            \boldsymbol{r}_{\mathrm{OT}} \cdot \boldsymbol{r}_{\mathrm{ST}}
        }{
            \| \boldsymbol{r}_{\mathrm{OT}} \|
            \| \boldsymbol{r}_{\mathrm{ST}} \|
        }
    \right)
    }
\end{equation}
where $\boldsymbol{r}_{\mathrm{OT}}$ is the relative position vector from the observer to the target, $\boldsymbol{r}_{\mathrm{ST}}$ is the relative position vector from the Sun to the target, and $p_{\mathrm{diff}} (\psi)$ is the diffuse phase angle function for a diffuse sphere, given by
\begin{equation}
    p_{\mathrm{diff}} (\psi) = \dfrac{2}{3\pi} \left[ \sin(\psi) + (\pi - \psi) \cos(\psi) \right]
\end{equation}

Evaluating the apparent magnitude of a target at all time steps over the mission duration yields a distribution function. The value of apparent magnitude $m(\alpha)$ in the distribution function is such that for $\alpha$\% of the time steps, the target is brighter than this value is extracted  and compared to a cutoff magnitude $\mco$. The target is then deemed visible if
\begin{equation}
    m (\alpha) \geq \mco
    \label{eq: visibility}
\end{equation}

Using the dynamics presented in the previous sections, candidate observers are placed in LPO families; examples of LPOs considered in this work are shown in Figure \ref{fig:lpo_families}.

Given a set of observers $\mathcal{O}$, a set of targets $\mathcal{T}$, the duration of the observation campaign $\Delta T$, and a time-step $\Delta t$, the simulation computes an observability metric $\Gamma$, which corresponds to the percentage of targets inside the grid that are deemed visible by (\ref{eq: visibility}), using Algorithm \ref{alg:simulator_base}.
\begin{algorithm}[h!]
\caption{Computation of observability metric for a given architecture}
\label{alg:simulator_base}
\begin{algorithmic}
\Require Observer set $\mathcal{O}$, target set $\mathcal{T}$, observation campaign duration $\Delta T$, time-step $\Delta t$
\State $N = \Delta T / \Delta t$
\For{$i = 1$ to $N$}  \Comment{for each time-step}
    \For{$o$ in $\mathcal{O}$}   \Comment{for each observer}
        \For{$t$ in $\mathcal{T}$}  \Comment{for each target}
            \State Update position of observer $o$
            \State Compute visible magnitude of $t$ from $o$
        \EndFor
    \EndFor
\EndFor
\State Compute $\Gamma$ as average number of visible targets over $N$ time-steps
\end{algorithmic}
\end{algorithm}

In practice, the algorithm is implemented via vectorization to avoid performance degradation due to the nested for-loops.

\section{Optimization of Cislunar Monitoring Architecture}
\label{sec:optim_hgga}
Leveraging the simulation environment for cislunar SDA, we consider a multiobjective optimization problem that simultaneously seeks to maximize the coverage and minimize the operational cost of the SDA architecture. 
First, this section introduces the optimization problem that is to be solved. 
Then, we introduce the HGGA, which enables the optimization of the SDA architecture without prescribing a constant number of observers a priori. 

\subsection{Optimization Problem Formulation}
The multiobjective optimization problem is formulated as follows:
\begin{equation}
    \min_{\mathbf{X}} \left( -\Gamma ,\,\, \sum_{j=1}^n i_j \nu_j , \,\, N_{\mathrm{spacecraft}} \right)
    \label{eq:objectives}
\end{equation}
Here, the objectives are to maximize the coverage, minimize the total stability index of all LPOs used, and minimize the total number of spacecraft.

The decision vector $\mathbf{X}$ consists of a list of observers:

\begin{equation}
    [\mathrm{Observer}_1,...,\, \mathrm{Observer}_{N_{\mathrm{spacecraft}}}]
\end{equation}

where an observer is defined by the 3 following variables:

\begin{equation}
    [\mathrm{Orbit\,Family},  P, \phi]
\end{equation}

and orbit families considered are

\begin{equation}
    [\mathrm{L1\,\,Halo_S, L2\,\,Halo_N, DRO, L2\,\,Lyap, L1\,\,Lyap}]
\end{equation}

A more detailed description of the variables of this problem is given in Table \ref{tab:variables}. Design variables that are traded off in this work are the total number of satellites, their respective allocated orbit family, their orbit period, and the initial phase angle along the orbit.
We can further decompose the variables into two distinct categories, discrete variables which are used to represent the orbit families. The rest of the variables will be treated as continuous variables. 

\begin{table}[h]
\begin{center}
\caption{Variables for the Cislunar SDA problem}
\begin{tabular}{@{}llll@{}}
\toprule
Discrete Variables & Continuous Variables  \\
\midrule
Orbit Family: $L1\,Halo_S^1,\cdots,L1\, Halo_S^{i_1}$  & Orbit Period: $P_{L1\, Halo_S^{i_1}},\cdots,P_{L1\,Halo_S^{i_1}}$ \\
\hspace{2cm} \vdots \hspace{1.4cm} \vdots  & \hspace{2cm} \vdots \hspace{1.5cm} \vdots \\       
\hspace{1.7cm} $L1\,Lyap^1,\cdots,L1\,Lyap^{i_n}$ & \hspace{1.7cm} $P_{L1\,Lyap^1},\cdots,P_{L1\, Lyap^{i_n}}$  \\
\\
& Phase Angle: $\phi_{L1\, Halo_S^{i_1}},\cdots,\phi_{L1\,Halo_S^{i_1}}$\\
& \hspace{2cm} \vdots \hspace{1.5cm} \vdots \\
& \hspace{1.7cm} $\phi_{L1\, Lyap^1},\cdots,\phi_{L1\,Lyap^{i_n}}$\\
\botrule
\end{tabular}
\label{tab:variables}
\end{center}
\end{table}

As a consequence of this formulation, the total number of spacecraft is an 
intrinsic variable comprised in the orbit family variable as follows.
\begin{equation}
    N_{\mathrm{spacecraft}} = \sum\limits_{j=1}^{n} i_j \,\,\,\,  (n = 5)
\end{equation}

\subsection{Hidden-Genes Genetic Algorithm}
\label{sec:hgga}

 Given the above formulation, discrete variable values are directly influencing the number of continuous variables, creating what's called a Variables-Size Design Space (VSDS) problem\cite{Gad2011,Abdelkhalik2018}. Traditional algorithms are not suited for the search of this type of design space.\

A relatively novel type of optimization algorithm, Hidden-Genes Genetic Algorithm (HGGA)\cite{Gad2011,Abdelkhalik2016,Abdelkhalik2018,Ellithy2022} allows in this work for the integer variable to enter as a design variable thus exploring the VSDS.

Studies on metaheuristic algorithms with variable-length chromosomes go back to the late 1990s \cite{Kajitani1996,Kim2005}. 
Specifically, the hidden-genes genetic algorithm (HGGA) have been recently investigated to use in the Multi Gravity Assist trajectory optimization problem tackling the complexity of assessing the optimal number of planets and the sequence of flybys.

\begin{figure}[ht!]
    \centering
    \includegraphics[width=1\textwidth]{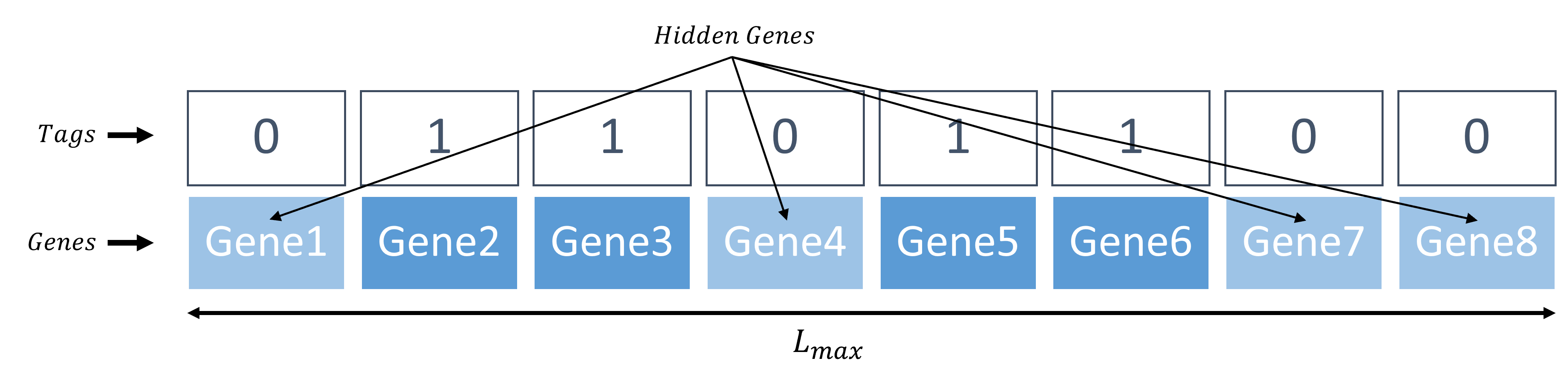}
    \caption{Chromosome with hidden genes \cite{Abdelkhalik2018}}
    \label{fig:Hidden_Gene}
\end{figure}

The HGGA method considers a fixed length of chromosome $L_{max}$, which contains the maximum number of design variables. In this case the maximum length would correspond to two times the total number of spacecraft given that each spacecraft is assigned two variables. As seen on Figure \ref{fig:Hidden_Gene}, a tag, associated to each gene, assesses if the gene is active or hidden. Hidden genes are not considered in the evaluation of the objective function.
Chromosomes with lower number of spacecraft are designed by switching genes to hidden genes in order to have the right number of active design variables that will be used to calculate the objectives to minimize.

In practice, tags evolve using stochastic processes. Some mechanisms can be found in \cite{Abdelkhalik2018}, while this works uses classical probability-based one-point crossover and bit-flip mutation operations for the tags. It is important to note that those operations are independent from the genes' mutation and crossover operations that take place in the regular genetic algorithm as can be seen in \ref{alg:HGGA}.

The multi-objective HGGA in this study is built upon the Non-dominated Sorting Genetic Algorithm II (NSGA-II) multi-objective genetic algorithm \cite{Deb2002}, using simulated binary crossover and polynomial mutations, akin to the implementation by Ellithy et al \cite{Ellithy2022}. 
Each operation on the genes and the tags is performed sequentially starting with the genes when creating a new generation of chromosomes, following Algorithm \ref{alg:HGGA}. 

\begin{algorithm}[h!]
\caption{Overview of the Hidden-Genes NSGA-II crossover and mutation mechanisms}
\label{alg:HGGA}
\begin{algorithmic}
\Require Parent population $\mathcal{P}$
\State Children population $\mathcal{C}$ = [ ] \Comment{Initialize list of children}
\While{Number of children $\neq$ Number of parents}  
    \State Select two parents randomly   
    \If{random(0,1) $>$ Gene Crossover Probability}  
        \State Perform gene crossover
    \EndIf
    \If{random(0,1) $>$ Tags Crossover Probability}  
        \State Perform tags crossover
    \EndIf
    \If{random(0,1) $>$ Gene Mutation Probability}  
        \State Perform gene mutation
    \EndIf
    \If{random(0,1) $>$ Tags Mutation Probability}  
        \State Perform tags mutation
    \EndIf
    \For{Both children}
        \For{Genes in children chromosome}
            \If{Tag associated to gene $\neq$ 0}
                \State Use gene in objective function evaluation
            \EndIf
        \EndFor
    \EndFor
    \State Add both children to $\mathcal{C}$
\EndWhile
\State Return $\mathcal{C}$
\end{algorithmic}
\end{algorithm}

\section{Numerical Results}\label{sec:numerical_results}
The proposed Cislulnar SDA simulation framework is solved using the HGGA.   
Table \ref{tab:problem_parameters} presents the parameters used for the simulation and the optimization. 
We first conduct parametric studies for multiple values of $\mco$ then analyze some of the best architectures extracted from the design space. The $m_{\mathrm{co}}$ limits were picked based on two reasons, the upper limit was chosen as the first apparent magnitude where literature's architectures' observability begin to decrease \cite{Holzinger2021_report}. The lower limit is the highest limiting magnitude attainable by state-of-the-art Cubesats telescopes for which the apparent magnitude usually ranges from 12 to 15 \cite{ashcraft2021versatile,aperture_optical_science_2021}.

\begin{table}[]
    \centering
    \caption{Cislunar SDA Simulation and Optimization Parameters}
    \begin{tabular}{@{}lll@{}}
    \toprule
    Category  & Parameter & Value \\ \midrule
    \multirow{9}{*}{Simulation} 
    & Initial epoch     &  2024 MAR 25 07:12:36.633 \\
    & Campaign duration $\Delta T$, day     &   1825    \\
    & Time-step $\Delta t$, day    &   2    \\
    & Number of targets &   1107        \\
    & Target diameter, $\mathrm{m}$   &   $1$  \\
    & Target specular reflectance $\aspec$ &   $0$   \\
    & Target diffuse reflectance $\adiff$ &    $0.2$   \\
    & Cut-off magnitude $\mco$ &   18, 17, 16, 15    \\
    & Fractional visibility requirement, $\alpha$  & 0.9   \\
    \hline 
    \multirow{8}{*}{Optimizer} 
    & Number of generations &  100     \\
    & Chromosomes per generation &  100 \\
    & Maximum chromosome length, $L_{max}$ & 30\\
    & Gene crossover probability &  0.95    \\
    & Distribution index for gene crossover &  10    \\
    & Gene mutation probability &  1/15   \\
    & Distribution index for gene mutation &  50    \\
    & Tag crossover probability &  0.8    \\
    & Tag mutation probability &  0.2    \\
    \bottomrule
    \end{tabular}
    \label{tab:problem_parameters}
\end{table}

\subsection{Overview of Pareto Front}
\begin{figure}[ht!]
    \centering
    \includegraphics[width=1\linewidth]{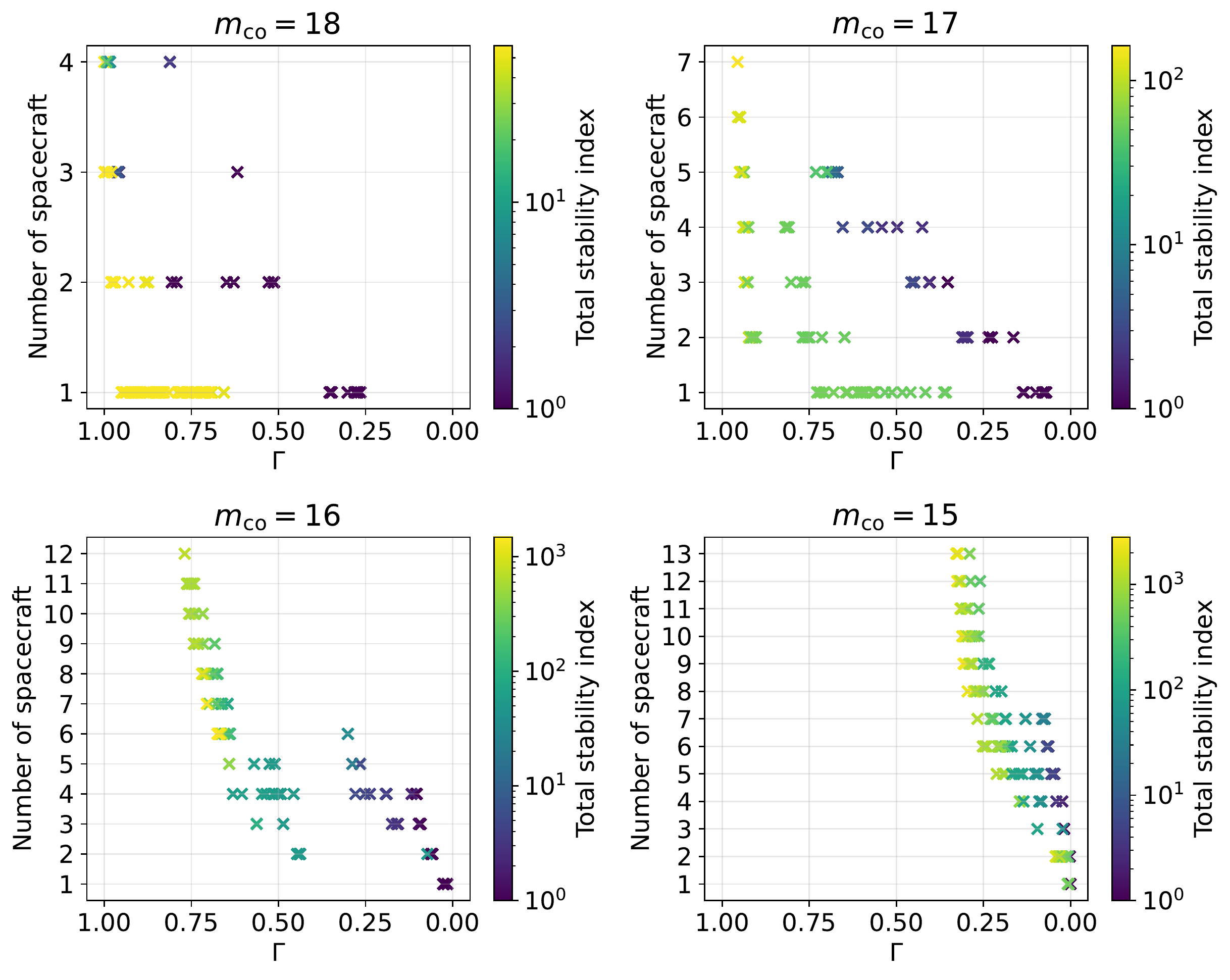}
    \caption{Pareto solutions of SDA architectures for $\mco$ between $18$ and $15$}
    \label{fig:pareto_combine}
\end{figure}

 As aforementioned, it is known there exists a gap in single spacecraft observability access \cite{Holzinger2021_report}, particularly for low apparent magnitudes. Figure \ref{fig:pareto_combine} shows that below a cutoff magnitude of 18, performances start degrading and $\Gamma$ drops to 0\% for magnitudes of 16 and below, when considering a single LPO.

\begin{figure}[ht!]
    \centering
    \includegraphics[width=0.8\linewidth]{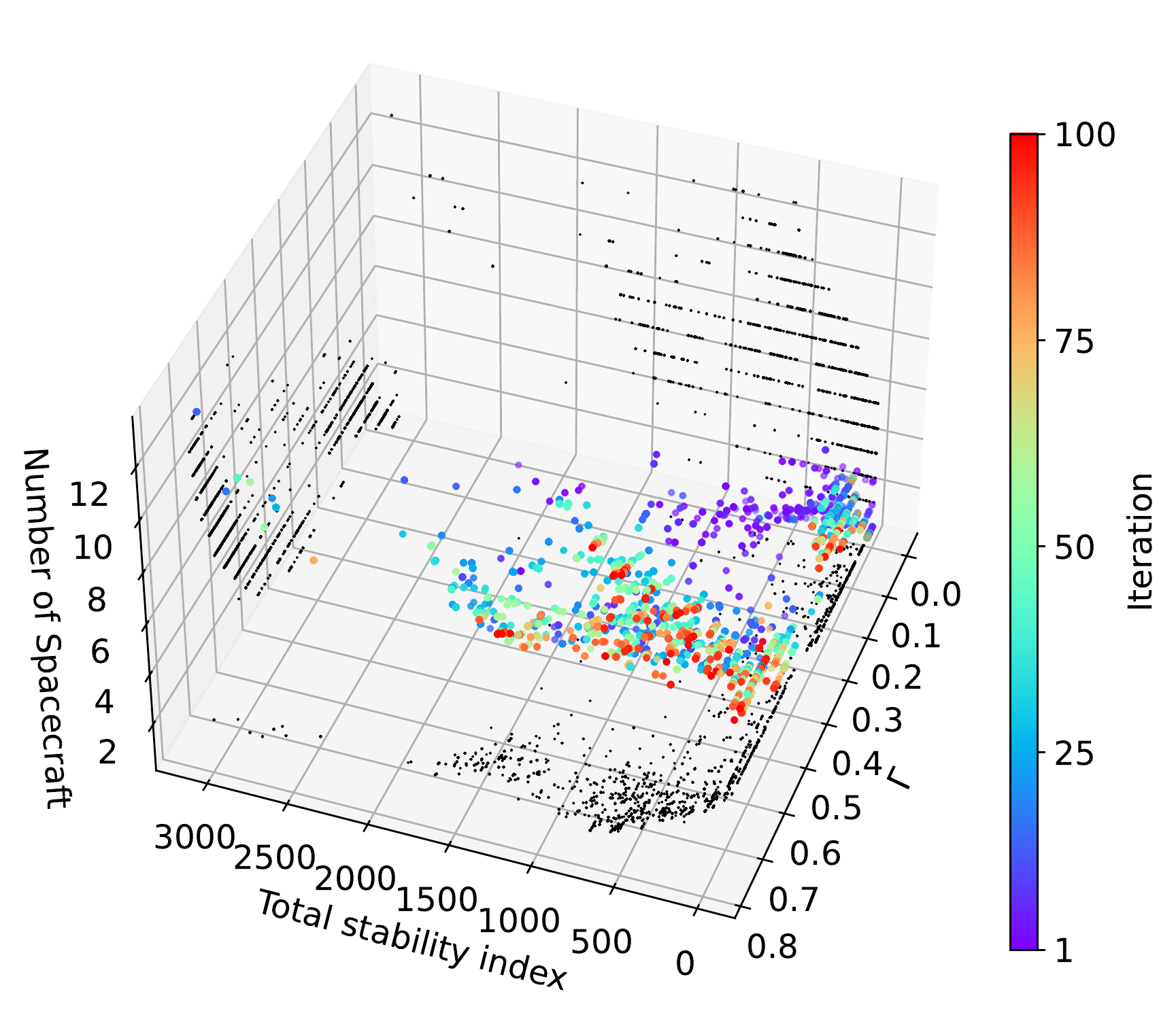}
    \caption{Evolution of the Pareto front for 3 objectives optimization with HGGA, $m_{\mathrm{co}}$ = 16. Black dots correspond to the projection of each solution across the three 2D planes.}
    \label{fig:pareto_evolve}
\end{figure}

The multi-spacecraft architectures in this work aim at filling this gap. 
The multi-objective HGGA extracts the Pareto front for the given objectives from all the possible solutions. 
As an example, Figure \ref{fig:pareto_evolve} presents the evolution of this Pareto front along the 100 iterations of the algorithm for the $\mco = 16$ case, underlining the apparent competitiveness between the stability index and $\Gamma$ objectives. 
It is also possible to note the nature of the optimization problem with respect to the three objectives; the projections indicate the optimizer is able to reduce both the total stability index and $\Gamma$ while maintaining the same number of spacecraft. 

As previously mentioned, for $m_{\mathrm{co}}$ of 16 and 15, the single spacecraft solutions only minimize the stability index metric.

The set of nondominated solutions for each $m_{\mathrm{co}}$ comprises many points. Some patterns are nonetheless visible.
For $m_{\mathrm{co}}$ = 18, single spacecraft solutions are already performing well, and adding new observers shifts the $\Gamma$ towards better solutions without greatly modifying the overall stability index. This shift continues as the number of spacecraft increases but full observability is quickly attained and the nondominated set, for the larger numbers of observers, therefore, contains $\Gamma=1$ solutions with lower stability indices.

Looking at the $m_{\mathrm{co}}$ = 17 plot, the shift is visible again.
An observability of around 95\% is attained for 2 observers, representing an improvement of 24\% compared to single spacecraft architectures. 
The gap to reach full observability is almost filled again while being larger. However, architectures seem unable to attain full observability. This is due to a certain percentage of points of the grid of targets never reaching an apparent magnitude of 17 since some of those points are simply too far away to be visible to the observers. This phenomenon becomes even more visible for lower cutoff magnitudes.

Figure \ref{fig:pareto_combine} shows, for $m_{\mathrm{co}}$ = 16, a sizeable augmentation of the $\Gamma$ between 1 and 2 observers, going from 1\% observability to approximately 44\%. This improvement is also seen from 2 to 4 observers, increasing to 63\%. Again, the flattening of the Pareto front, described for $m_{\mathrm{co}}$ = 17, is visible. In fact, for a higher number of spacecraft, the gain in $\Gamma$ is reduced, while the stability indices rise rapidly. 

Finally, with $m_{\mathrm{co}}$ = 15, increasing the number of spacecraft has limited improvement on $\Gamma$. 
The maximum $\Gamma$ attained is 33\% and the total stability index is 2267.48, highlighting fairly unstable orbits.
This drop in $\Gamma$ between $m_{\mathrm{co}}$ of 16 and 15 is once again due to the percentage of targets that never exceed an apparent magnitude of 15 given the observers position becomes too important, in particular in the near-Earth areas of the grid where the LPO become too distant to perceive the target spacecraft. 

The best results in terms of $\Gamma$ for each $m_{\mathrm{co}}$ and the related number of spacecraft and stability indices are given in Table \ref{tab:results}.

\begin{table}[]
    \centering
    \caption{Cislunar SDA results}
    \begin{tabular}{@{}llllllll@{}}
    \toprule

    & \multicolumn{3}{l}{Single Spacecraft} & \multicolumn{4}{l}{Multi Spacecraft}\\
    \cline{2-4}
    \cline{5-8}
    $m_{\mathrm{co}}$  &  $N_{\mathrm{spacecraft}}$ & $\Gamma$ & Total $\nu$ & $N_{\mathrm{spacecraft}}$ & $\Gamma$ & Total $\nu$ & Average $\nu$\\ \midrule
 
    \multirow{2}{*}{18}  & \multirow{2}{*}{1} & \multirow{2}{*}{0.95}   &  \multirow{2}{*}{55.18}  & 2 & 0.98 & 56.29 & 28.14\\
     &  &     &   & 3 & 1.0 &  57.33 & 19.11\\
    \hline
    \multirow{3}{*}{17}  & \multirow{3}{*}{1} & \multirow{3}{*}{0.73 }   &  \multirow{3}{*}{55.18}  & 2 & 0.92 & 100.17 & 50.08\\
     &  &     &    & 3 & 0.94 &  119.66 & 39.89\\
    &  &     &    & 7 & 0.96 & 163.25 & 23.32\\
    \hline 
    \multirow{6}{*}{16}  & \multirow{6}{*}{1} & \multirow{6}{*}{0.03 }   &  \multirow{6}{*}{1.00}  & 2 & 0.44 & 51.11 & 25.55\\
     &  &     &    & 4 & 0.63 &  58.06 & 14.52\\
    &  &     &    & 6 & 0.67 &  1419.99 & 236.67\\
    &  &     &    & 8 & 0.72 & 1095.21 & 136.90\\
    &  &     &    & 10 & 0.76 & 604.79 & 60.48\\
    &  &     &    & 12 & 0.77 & 720.87 & 60.07\\
    \hline 
    \multirow{6}{*}{15}  & \multirow{6}{*}{1} & \multirow{6}{*}{0.01 }   &  \multirow{6}{*}{536.80}  & 2 & 0.04 & 1790.04 & 895.02\\
    &  &     &    & 4 & 0.15 &  701.61 & 175.40\\
    &  &     &    & 6 & 0.25 &  1137.78 & 189.63\\
    &  &     &   & 8 & 0.30 & 2261.87 & 282.73\\
    &  &     &    & 10 & 0.31 & 2181.53 & 218.15\\
    &  &     &   & 13 & 0.33 & 2267.48 & 174.42 \\

    \bottomrule
    \end{tabular}
    \label{tab:results}
\end{table}

\subsection{Analysis of Architectures}
\begin{figure}[ht!]
    \centering
    \includegraphics[width=1\linewidth]{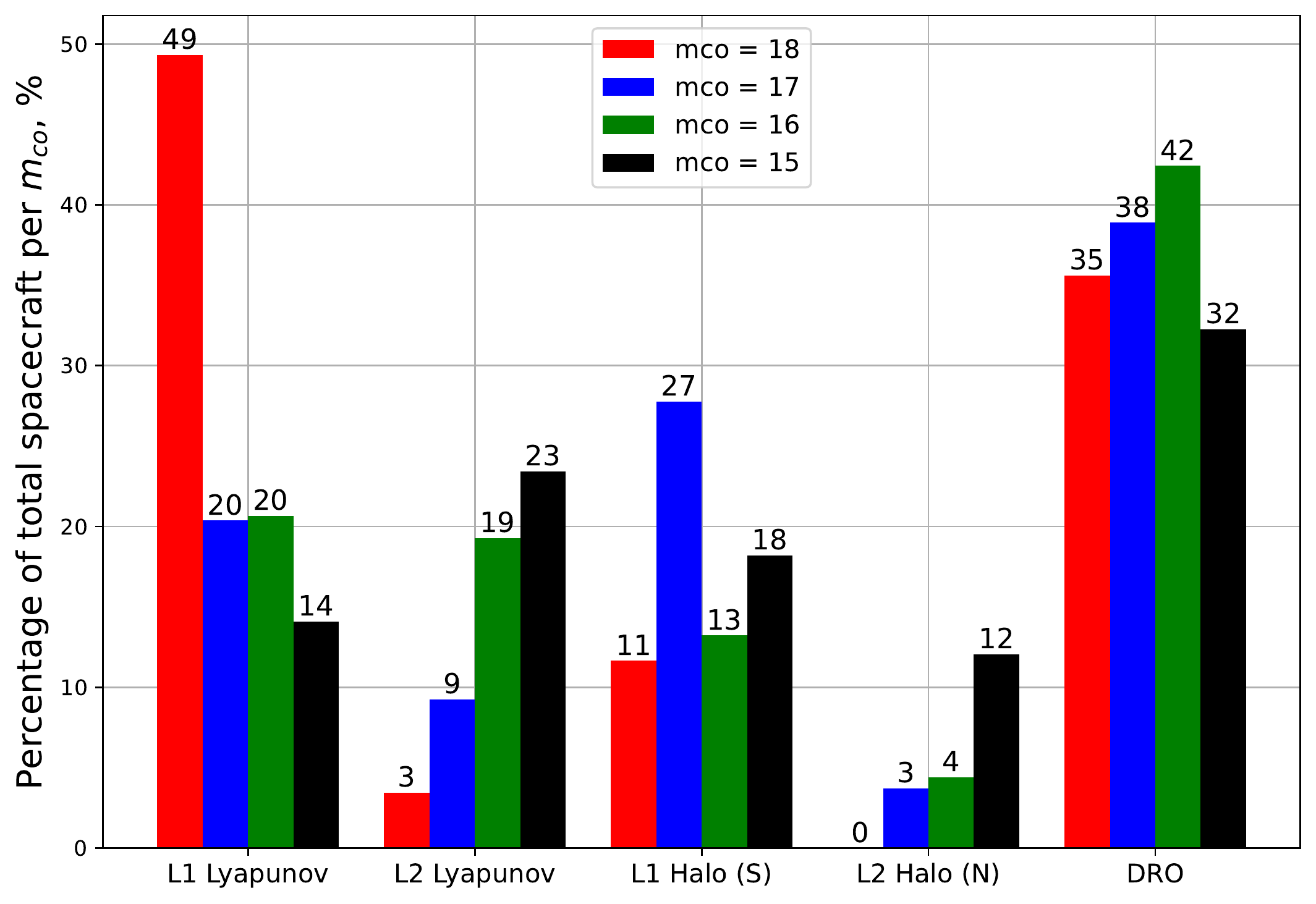}
    \caption{Relative distribution of spacecraft among orbit families for different $m_{co}$}
    \label{fig: Distribution}
\end{figure}

Among all the architectures selected by the HGGA, an evolution in the allocation of spacecraft to orbit families is clearly visible when varying the $m_{\mathrm{co}}$. Figure \ref{fig: Distribution} shows the number of times a particular orbit is utilized by the Pareto front solutions for all values of $m_{\mathrm{co}}$ considered. 

Overall, apart from an $m_{\mathrm{co}}$ value of 18, the L1 Lyapunov family is used by any value of $m_{\mathrm{co}}$ at roughly equal frequencies. This can be explained by the family being at the center of the target grid, thus offering an effective position for covering large portions of the region of interest. For $m_{\mathrm{co}}$ = 18,  single spacecraft architectures are already nearly optimized using a single L1 Lyapunov orbit and closing the gap only requires DROs, which are the most stable option.
Meanwhile, it is also possible to observe that as the $m_{\mathrm{co}}$ is reduced, the architecture begins to also require an observer at the vicinity of L2 dedicated to targets on the far end. 
Also, the low $\nu$ of DROs make them a particularly attractive family vis-a-vis the second objective of the problem from \eqref{eq:objectives}, resulting also in their relative frequent use. 

For each value of $m_{\mathrm{co}}$, it is also possible to observe intuitive variations of family allocations as the number of spacecraft increases.
As  $m_{\mathrm{co}}= 16$ offers the largest changes in $\Gamma$ when adding new spacecraft and overall the most important increases in performance, we focus further on analyzing different architectures for this particular cutoff magnitude. We select architectures encompassing various total number of spacecrafts, while having reasonable total $\nu$. For those reasons, 2-, 3-, 5- and 7-spacecraft architectures with the highest $\Gamma$ are selected from the Pareto front, at the exception of the 7-spacecraft solution where a more stable one is selected with a slightly lower observability metric.

Table \ref{tab:results_Period} summarizes the information of the LPOs used by these architectures. 

\begin{figure}[ht!]
    \centering
    \includegraphics[width=1\linewidth]{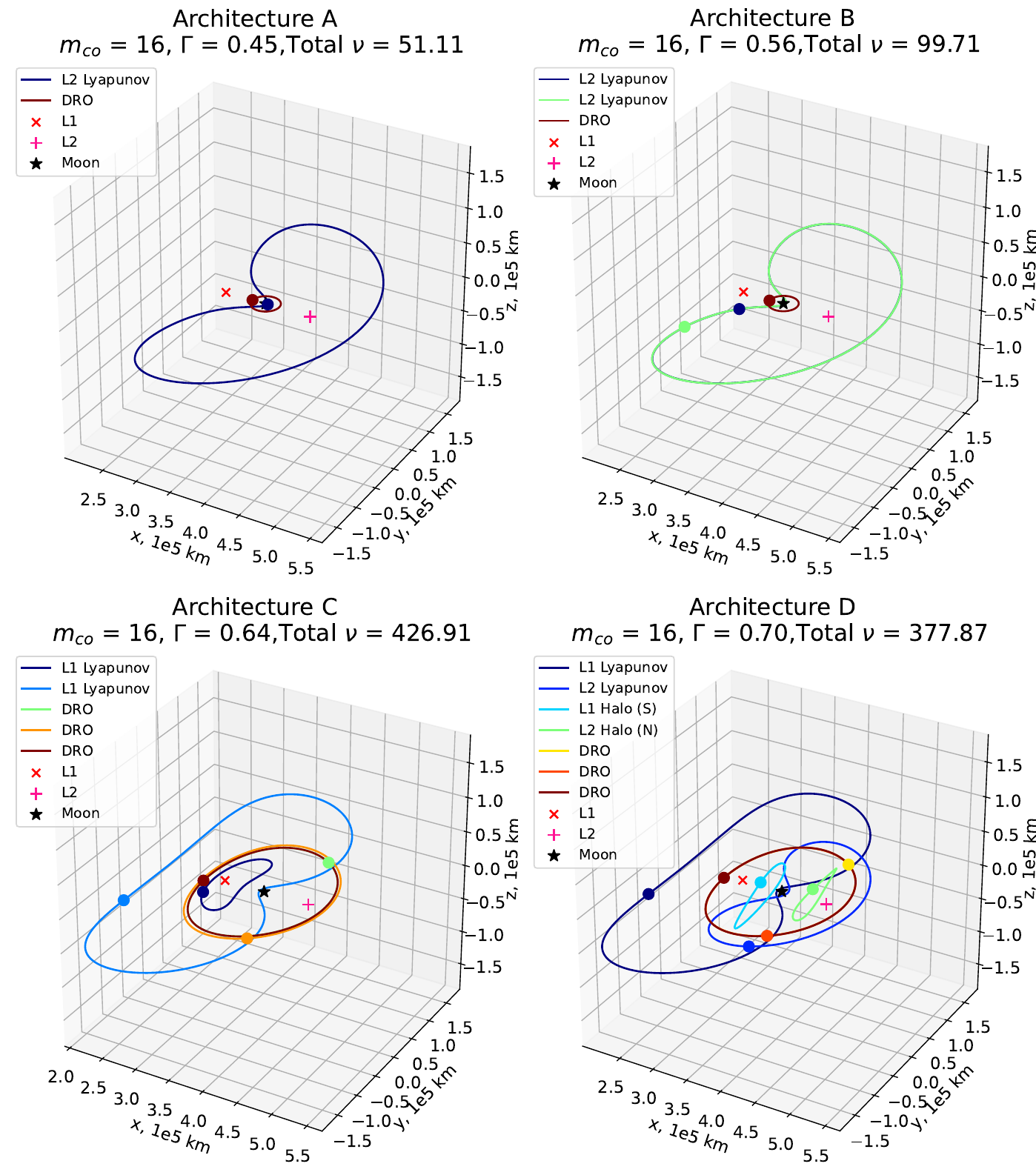}
    \caption{Example Pareto solutions with 2-, 3-, 5- and 7-spacecraft architectures for $m_{\mathrm{co}}$  = 16}
    \label{fig: Combined Architectures}
\end{figure}

Architecture A and B on Figure \ref{fig: Combined Architectures} shows the low number of spacecraft solutions that achieved the highest $\Gamma$.
Directly, as opposed to higher $m_{\mathrm{co}}$ minimal spacecraft number architectures, the L1 Lyapunov orbits have been replaced by L2 Lyapunov orbit which confirms the trend discussed previously.

Another interesting thing to notice is that from 2 to 3 spacecraft, the orbits family and orbits periods variables are close to unchanged and the optimized solution simply is derived by the optimizer using an offset of the initial phasing of the L2 Lyapunov orbit observers while enabling an increase in approximately 24\% of the observability.

Also, in both cases, the L2 Lyapunov orbits use 1:1 synodic resonant orbits, combined with an initial phase angle close or equal to zero, hence aligned with the original sun-earth-moon positions as discussed in Section \ref{sec:simulator}. Resonance with the synodic period has been proven to provide particularly beneficial light conditions throughout the mission horizon \cite{Holzinger2021_report}.

When augmenting the number of spacecraft per architecture, the variability in the combinations of orbit families assigned to spacecraft augments as well. 

This is particularly noticeable for architectures C and D on Figure \ref{fig: Combined Architectures} where the 5-spacecraft architecture already differs considerably from the low number of spacecraft combinations in terms of orbit families considered. 
Although the architecture C only considers two orbit families, we can notice a 120° off-set between the 3 DRO observers that allows for continuous monitoring, even in the varying light conditions, as there are higher chances that one of the three will have good alignment with regards to the sun. This can be seen as similar mechanism as the resonant period, adapted to the DROs for which resonant orbits are not particularly suited, given the great variability in size for adjacent orbit periods.
Again, it is possible to observe 2:1 and 1:1 resonance with close to 0° initial phasing for the L1 Lyapunov families. This architecture on the other hand presents poor stability characteristics, especially when compared to other solutions with more spacecraft.

The 7-observers configuration offers a great geometrical diversity in terms of orbit, while resulting in a lower total $\nu$ and an even more reduced average $\nu$ than some of the other combinations considering fewer spacecraft.
The particular 3 DROs configuration has been preserved for lower number of spacecraft solutions as well as the 1:1 resonant L1 Lyapunov.
Added to this is an approximate 2:1 south L2 Halo. The last two orbits don't exhibit any particular geometry characteristics of their own but are efficient when associated to the rest of the orbits.

\begin{table}[ht!]
    \centering
    \caption{Period and initial phase angle from selected solutions}
\begin{tabular}{@{}lllll@{}}
\toprule
Solution           & Orbit Family & Orbital period $P$, days & \begin{tabular}[c]{@{}l@{}} Fraction of \\ synodic period \end{tabular}
& Phase angle $\phi$, deg \\ \midrule
\multirow{2}{*}{A} & L2 Lyapunov  & 29.54         & 1.00                & 0.00                     \\
                   & DRO          & 2.54          & 0.09                & 0.00                     \\
                   \hline
\multirow{3}{*}{B} & L2 Lyapunov  & 29.48         & 1.00                & 347.44                   \\
                   & L2 Lyapunov  & 29.50         & 1.00                & 322.05                   \\
                   & DRO          & 2.51          & 0.09                & 353.20                   \\
                   \hline
\multirow{5}{*}{C} & L1 Lyapunov  & 14.78         & 0.50                & 328.72                   \\
                   & L1 Lyapunov  & 29.48         & 1.00                & 341.01                   \\
                   & DRO          & 16.55         & 0.56                & 117.14                   \\
                   & DRO          & 17.14         & 0.58                & 240.06                   \\
                   & DRO          & 16.55         & 0.56                & 353.83                   \\
                   \hline
\multirow{7}{*}{D} & L1 Lyapunov  & 29.47         & 1.00                & 344.40                   \\
                   & L2 Lyapunov  & 23.03         & 0.78                & 245.00                   \\
                   & L1 Halo (S)  & 11.62         & 0.39                & 204.20                   \\
                   & L2 Halo (N)  & 14.32         & 0.49                & 226.08                   \\
                   & DRO          & 16.47         & 0.56                & 120.85                   \\
                   & DRO          & 16.47         & 0.56                & 239.79                   \\
                   & DRO          & 16.50         & 0.56                & 357.07                  \\
\hline
\end{tabular}
\label{tab:results_Period}
\end{table}

As can be observed, when apparent magnitude conditions become more strict, diversity in the allocations of observers begins to be necessary to achieve performing $\Gamma$.
Performing L1 Lyapunov combinations are still possible at the cost of a higher overall stability index.

\section{Conclusion}\label{sec:conclusion}
In this work, we have studied the design optimization of multispacecraft cislunar SDA architectures. The architecture considered is tasked with 
a static, grid-based region of interest observation based on a metric $\Gamma$ defined as the percentage of this region of interest visible over a fraction $\alpha$. Two other metrics, the total stability index and the total number of spacecraft are introduced to compete with the first objective and allowed us to obtain a nondominated set of architectures solutions. Given the variability in the size of the design space, due to the fact that the number of spacecraft and the orbit family allocation both enter as design variables, a Hidden-Genes Non-dominated Sorting Genetic Algorithm II is implemented. 

Solutions obtained aim at filling some of the gaps in space domain awareness monitoring for lower value of limiting magnitude, allowing for less advanced, and hence cheaper space telescope constellations to complete the mission.
Results showed that it was possible to reduce the limiting magnitude of the monitoring structures, greatly, increasing the $\Gamma$ for a reasonable number of spacecraft and total $\nu$. In particular, diversity in the choice of orbit families is key to achieving better observability, as shown in Figure \ref{fig: Distribution}. 
Another common solution extracted by the optimizer is the use of a small phase difference between observers from the same orbit family with relatively similar periods, in order to achieve almost constant ideal light conditions on the whole cislunar space.

The framework and optimization formulation utilized in this work are very flexible, allowing for example to add new types of observers without changing the formulation. These new optical structures could be Low Lunar Orbits, Lunar Surface Telescopes or even Earth-orbiting observers such as geosynchronous orbits \cite{Klonowski2022} or Highly Eccentric Orbit (HEO) \cite{Cunio2020} which have been proven to be particularly useful for cislunar trajectory monitoring. 
It is also possible to change the grid of targets, creating denser areas in potential congestion points or along the most likely trajectories for lunar transfer orbits.
Finally, the introduction of more concrete cost metrics such as transfer costs or station-keeping may become interesting in the coming years as cislunar space domain awareness carry on its rapid development.

\bmhead{Acknowledgments}

This work has been funded by the Georgia Tech Research Institute (GTRI)'s Independent Research and Development (IRAD) program.

\bibliography{sn-bibliography.bib}  


\end{document}